# Selective and Fast Plasmon-Assisted Photo-Heating of Nanomagnets; A New Route for Opto-Activated Nanomagnetic Logic and Artificial Spin Systems

*Matteo Pancaldi[1,†], Naëmi Leo[1], and Paolo Vavassori[1,2]\**

[1]CIC nanoGUNE, Donostia-San Sebastian, E-20018, Spain; [2]IKERBASQUE, Basque Foundation for Science, Bilbao, 48013, Spain

[†]Present address: Department of Physics, Stockholm University, 106 91 Stockholm, Sweden.

\*E-mail: p.vavassori@nanogune.eu

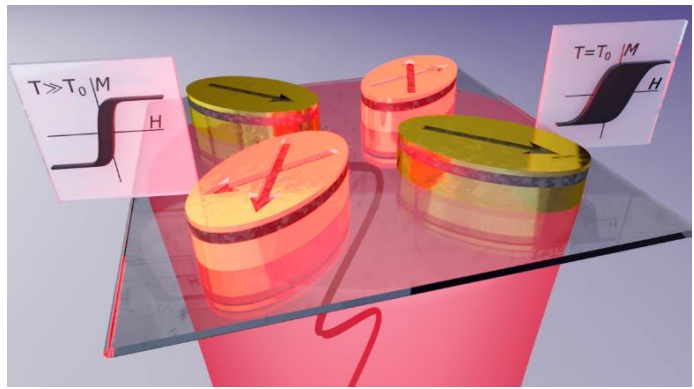

Thermal relaxation of nanoscale magnetic islands, mimicking Ising macrospins, is indispensable for studies of geometrically frustrated artificial spin systems and low-energy nanomagnetic computation. Currently-used heating schemes based on contact to a thermal reservoir, however, lack the speed and spatial selectivity required for the implementation in technological applications. Applying a hybrid approach by combining a plasmonic nanoheater with a magnetic element, in this work we establish the robust and reliable control of local temperatures in nanomagnetic arrays by contactless optical means. Plasmon-assisted photo-heating allows for temperature increases of up to several hundred Kelvins, which lead to thermally-activated moment reversals and a pronounced reduction of the magnetic coercive field. Furthermore, the polarization-dependent absorption cross section of elongated plasmonic elements enables sublattice-specific heating on sub-nanosecond time scales. Using optical degrees of freedom, *i.e.* focal position, polarization, power, and pulse length, thermoplasmonic heating of nanomagnets offers itself for the use in flexible, fast, spatially-, and element-selective thermalization for functional magnetic metamaterials.

# INTRODUCTION

Networks of nanomagnets interacting via magneto-static interactions are key metamaterials for magnetic storage devices,[1] for low-power information processing,[2-4] and as model system to create and quantify collective phenomena.[5-7] In recent years, so-called artificial spin systems – lithographically-defined two-dimensional arrays of elongated single-domain nanomagnets that behave like giant magnetic moments – with geometrically-frustrated lattices of nanomagnets reminiscent of spin-ice materials found in nature, have been used to quantify the statistics of low-energy states,[8-10] phase transitions,[11,12] and the properties of dynamic excitations such as magnetic monopoles.[13,14]

The study of the emergence and dynamics of collective equilibrium properties in the artificial magnetic metamaterials requires exploration of the systems' phase space, which is determined by the behavior of the single elements and the mutual magnetostatic interactions. Over the years, different schemes have been employed to drive artificial spin systems to an equilibrium state, including the study of as-grown samples,[5,8,15] demagnetization protocols,[16] heat treatments,[9,10] or the fabrication of superparamagnetic elements.[17,18] In addition, different experimental approaches – mostly based on large-scale facility techniques – allow to study the response of magnetic materials on different length- and time scales.[12,19,20]

Despite the substantial efforts to improve experimental characterization, as of today, the thermal excitation of artificial spin systems is achieved by thermal contact to a hot reservoir. This approach is energetically inefficient, spatially non-discriminative, and intrinsically slow, with a time scale of seconds to hours, making it difficult to reach a true equilibrium state in extended frustrated lattices. Furthermore, for implementation in devices of magnetic metamaterials, *e.g.* magnonic crystals[21,22] and magnetic logic circuits,[2,3,23,24] global heating lacks the spatial discrimination and speed required for integrated operation with CMOS technology. Therefore, the development of efficient low-power, non-invasive (*i.e.* contactless) heating which allows control over length scales and time scales of the thermal excitation of nanoscale magnetic metamaterials would enable deeper studies of equilibrium properties and emergent excitations in artificial spin systems, as well as open doors for the practical use in applications.

The emerging field of thermoplasmonics offers such an approach to fast and selective heating of nanostructures based on the excitation of localized surface plasmon resonances in nanostructures, usually made from gold. These collective electron oscillations at the metal-dielectric interface are induced by electromagnetic radiation[25,26] and the energy dissipation of the local currents lead to efficient, fast, and sizable temperature increase within the nanostructure,[27-29] which can even lead to ultrafast melting.[30]

Using lithographically-defined thin-film gold ellipses with typical thickness of several nanometers, and widths and length in the order of hundreds of nanometers, local temperature increases up to several hundred Kelvin can be easily achieved by the optical excitation of plasmonic resonances in the visible or near-infrared spectral region using common laser sources.[27,28,31] Such dimensions are comparable to those of single-domain nanomagnets commonly used for the study of artificial spin systems and nanomagnetic logic circuits. Therefore, a hybrid system combining a gold nanoantenna to support intense localized surface plasmon excitations for efficient thermoplasmonic heating with a magnetic element for the desired magnetic properties offers an attractive approach to achieve local and fast heating of interacting nanomagnetic systems.

Furthermore, for both plasmonic as well as nanomagnetic systems the elongated shape of the individual elliptical elements allow for additional functionality: In the case of plasmonics, the lateral dimensions of an elongated element determine the spectral position of the light-induced collective electronic excitations.[25,26] In the case of magnetism, the size and aspect ratio of the single-domain nanomagnetic element determine its magnetic anisotropy, which in turn determines the preferential direction of the net moment (*i.e.* the magnetic easy axis of the Ising macrospin), and how easily the net magnetization can be reversed by an applied magnetic field (*i.e.* the coercive field).[32,33] Due to the polarization-dependent excitation of plasmonic resonances in elliptical nanoislands, element-specific heating in an array of nanoantenna could be achieved, where, *e.g.* one sublattice shows a marked increase of temperature, whereas another – perpendicularly-oriented – sublattice retains the ambient background temperature. In contrast to conventional means, *i.e.* global heating, thermoplasmonic excitations offer additional degrees of freedom for light-controlled spatial- and element-selective heating of nanomagnets via variations of focal position, light polarization, and pump power, and thus allows for unique control of magnetic moment relaxation in magnetic metamaterials.

In this work, we apply this modular approach to demonstrate plasmon-assisted photo-heating of nanomagnets, combining gold nanoantennas for efficient photo-induced plasmonic heating with elements made from magnetic permalloy (Py, $Ni_{0.8}Fe_{0.2}$). We experimentally quantify the optical and magnetic properties of arrays of non-interacting hybrid elements as well as vertex-like assemblies, and present strategies to achieve efficient and selective control of the thermally-activated magnetic reversal by choice of focal point, pump power, light polarization, and pulse duration. These results establish plasmon-assisted heating for contemporary thermalization schemes in artificial spin systems or nanomagnetic circuits for computation.

RESULTS

To demonstrate the effect of plasmonic heating on the properties of nanomagnetic assemblies, we fabricated arrays of tri-layer Au-Py-Au elliptical nanoislands with lateral dimensions smaller than ≈200 nm on a transparent glass (Pyrex) substrate, as the schematic rendering in Fig. 1(a) shows (see Methods for further details on sample fabrication). The small size and elliptical shape of the magnetically-soft permalloy component stabilizes the magnetic single-domain state and promotes a magnetization reversal via coherent rotation, enabling to use these nanoislands as artificial macrospins for the study of artificial spin systems or nanomagnetic logic circuits.

After a systematic study using COMSOL Multiphysics simulations[34] and extinction spectra measurements of samples where the Au thickness was varied from 10 nm to 30 nm in steps of 5 nm, the thickness of the bottom Au layer was chosen to be 25 nm to achieve effective light-matter interaction in the desired spectral range. A 10 nm-thick Py film provides the desired magnetic properties, and a top layer of 5 nm Au prevents oxidization of the Py nanostructures. We consider a set of samples for which the ellipse islands share the same minor axis dimension $a_{\text{minor}} = 100$ nm and have increasing major axis length $a_{\text{major}} = AR \cdot a_{\text{minor}}$ with the aspect ratios $AR = 1.50, 1.75,$ and $2.00$. The islands are placed on a rectangular lattice with edge-to-edge distances of 230 nm, such that optical and magnetic interactions between the individual islands can be neglected. Scanning electron microscopy (SEM) images of the three arrays are shown in Figs. 1(b-d).

The hybrid gold-permalloy islands show plasmonic resonances in the visible/infrared spectral range, as shown in the extinction spectra, $1 - T(\lambda, \boldsymbol{E})$ where $T(\lambda, \boldsymbol{E})$ is the transmittance spectrum, shown in Figs. 1(e-g). Here, for incident light polarization along the ellipse major axis (thick lines), a pronounced evolution with increase of aspect ratio is observed:[25,31] First, the spectral position of the peak maximum shows a red shift, from 730 nm for $AR$ =1.50, Fig. 1(e), to 870 nm at $AR$ =2.00, Fig. 1(g). Second, with increasing aspect ratio, the peak height of the resonance is enhanced as well, with an increase of almost 30% from $AR$ =1.50 to $AR$ =2.00. In contrast, light which is polarized along the constant-width minor axis excites a plasmonic resonance at smaller wavelengths ~630 nm (thin lines), with no relevant change in peak position or peak height upon variation of the island aspect ratio.

In general, optical extinction spectra combine both the effects of absorption and scattering, with only the absorbed fraction of the light available for heat generation in the nanostructure. The alikeness of the extinction spectra $1 - T(\lambda, \boldsymbol{E})$, Figs. 1(e-g), and the simulated absorption cross section $\sigma_{abs}(\lambda, \boldsymbol{E})$, Figs. 1(h-k), indicated that – depending on wavelength and light polarization – the nanostructures will either absorb the power of the light field to a large extent, enabling efficient thermoplasmonic heating, or are almost transparent to the pump beam, and thus remain cool when illuminated.

To quantify this assumption, we measured magnetic hysteresis loops $\boldsymbol{M}(\boldsymbol{H})$ using the magneto-optical Kerr effect (MOKE) in a reflection geometry from the front of the sample, while simultaneously exciting the plasmonic gold structures by illumination through the back of the glass substrate, using a custom-built setup, as illustrated in Fig. S1 (see Methods for further details). In the chosen experimental geometry, the MOKE signal reflects the changes of the magnetization $\boldsymbol{M}(\boldsymbol{H})$ parallel to the magnetic field $\boldsymbol{H}$, which was applied along the long (magnetic easy) axis of the nanoislands, $i.e.$ resulting in the so-called easy-axis hysteresis loops shown in Fig. 2(a-c). The efficient heating with polarized light along the ellipse major axis supports thermally-activated magnetization reversal in the individual nanoislands. This leads to a reduction of the coercive field $H_C$, $i.e.$ where the islands' magnetization $\boldsymbol{M}(\boldsymbol{H})$ crosses its zero value. The value of $H_C$ depends on the shape anisotropy of the elliptical nanomagnet, given by its aspect ratio $AR$,[32,33] and the temperature-dependent saturation magnetization $M_S(T)$ of permalloy, which becomes non-magnetic above its Curie temperature $T_C$ =843 K. Due to the reduction of $M_S(T)$ upon heating, the experimentally-measured decrease of the coercive field $H_C(P_{\text{pump}})$ in dependence of the pump beam power $P_{\text{pump}}$ can be used for magnetic thermometry.

For each of the aspect ratios $AR$ =1.5, 1.75, and 2.00, MOKE hysteresis loops were measured with a pump beam polarised along the ellipse major axis at a wavelength of $\lambda_{\text{pump}}$=785 nm, where thermoplasmonic heating is expected to be most efficient for $AR$ =1.75, and varying power $P_{\text{pump}}$. The outermost loops (thick lines) in Figs. 2(a-c) correspond to ambient conditions, $i.e.$ $P_{\text{pump}} = 0$, and the initial coercive field $H_c(0)$, marked by circles, increase with ellipse aspect ratio $AR$ from 105 Oe for $AR$ =1.50 to 263 Oe for $AR$ =2.00, as expected from the variation in shape anisotropy and particle volume.[32,33] The succession of hysteresis loops, thin lines in Figs. 2(a-c), with increasing pump power $P_{\text{pump}}$ shows a continuous decrease of the coercive field $\Delta H_c^{\text{MOKE}}(P_{\text{pump}})$, as shown in Fig. 2(d-f). For $AR$ =1.75, for which the strongest light-matter interaction is expected at the pump wavelength of $\lambda_{\text{pump}}$=785 nm, the value of $H_c$ is almost reduced by 50% at maximum $P_{\text{pump}}$ of 60 mW, as highlighted by the shaded area in Figs. 2(b).

To relate the experimentally-observed decrease of $H_c^{\text{MOKE}}(P_{\text{pump}})$ with plasmonic pump power $P$ to a temperature increase of the sample, we employed micromagnetic simulations of the field-induced magnetization reversal to obtain values of $H_c^{\text{sim}}(T)$, varying the saturation magnetization $M_s(T)$ only (see Methods).[35] As shown in Figs. 2(d-f), the comparison of the coercive field values obtained by MOKE measurements $H_c^{MOKE}(P_{pump})$ (dots, bottom scale) and micromagnetic simulations $H_c^{sim}(T)$ (solid line, top scale) indicates a sizable temperature increase by thermoplamonic heating, *e.g.* ~320 K for the hybrid nanostructures with the aspect ratio of $AR=1.75$, Fig. 2(e). For the other two samples, $AR=1.50$ and $AR=2.00$, the expected temperature increase is ~220 K and ~240 K, respectively.

The relationship between $\Delta H_c^{MOKE}(P_{pump})$ and $\Delta H_c^{sim}(T)$ gives one estimate of the thermoplasmonic temperature increase based on the magnetic response of the samples. Another, independent, estimate based on the optical and structural properties of the nanoelements can be obtained by the following approximation of $\Delta T$ for nanostructures illuminated by a steady-state continuous-wave Gaussian beam of power $P_{pump}$ and full-width-half-maximum beam diameter $w_{FWHM}^{pump}$.[27,31,36]

$$\Delta T = \Delta T_{self} + \Delta T_{coll}$$
$$\approx \gamma_1 \frac{\sigma_{abs}(\lambda, \boldsymbol{E}) \cdot P_{pump}}{\kappa_{int} \cdot a_0 \cdot (w_{FWHM}^{pump})^2} + \gamma_2 \frac{\sigma_{abs}(\lambda, \boldsymbol{E}) \cdot P_{pump}}{\kappa_{int} \cdot S \cdot w_{FWHM}^{pump}} \left(1 - \gamma_3 \cdot \frac{\sqrt{S}}{w_{FWHM}^{pump}}\right). (1)$$

Here, $\sigma_{abs}(\lambda, \boldsymbol{E})$ is the wavelength- and polarization-dependent absorption cross section, as shown in Figs. 1(h-k), $\kappa_{int} = \frac{1}{2}(\kappa_{sub} + \kappa_{env})$ is the average thermal conductivity at the interface between substrate and surrounding medium, and the constants take the following values: $\gamma_1 = \frac{\ln 2}{\pi^2}$, $\gamma_2 = \sqrt{\frac{\ln 2}{4\pi}}$, and $\gamma_3 = \frac{4\sqrt{\ln 2}}{\pi}$. The first summand $\Delta T_{self}$ in Eq. (1) describes the local self-heating of an isolated nanostructure with an equivalence radius $a_0$, *i.e.* the radius of a sphere with the same volume as the illuminated nanoparticle. The second summand $\Delta T_{coll}$ describes the temperature increase due to the simultaneous excitation of several nanoantenna, as determined by the ratio between the unit cell area $S$ (respectively the nanoparticle surface density $S^{-1}$) and the beam diameter $w_{FWHM}^{pump}$. In this case, diffusion via the substrate of the heat generated in neighboring nanoislands leads to an overall increase of the temperature of the entire illuminated area.

Using Eq. (1), we can estimate the expected temperature increase of nanoislands with an aspect ratio of AR=1.75 due to plasmonic heating using the following values: $\kappa_{int}$=0.7 W m$^{-1}$ K$^{-1}$ (*i.e.* the mean value between the Pyrex glass substrate and vacuum; see Methods), $a_0$=50 nm, S=(330·405) nm$^2$, illuminated by a beam with diameter $w_{FWHM}^{pump}$=25 µm and power $P_{pump}$=60 mW. The values of $w_{FWHM}^{pump}$ and $P_{pump}$ correspond to those used for the MOKE experiments shown in Fig. 2, and under these conditions the collective contribution $\Delta T_{coll}$ dominates over the localized response $\Delta T_{self}$ by a factor of ≈30. With the absorption cross section $\sigma_{abs}(785\,nm, \boldsymbol{E}_{major})$ =5.0·10$^4$ nm$^2$, obtained by COMSOL simulations for the islands with the aspect ratio $AR=1.75$ in Fig. 1(i), a temperature increase of $\Delta T=306$ K is calculated. For $AR=1.50$ and $AR=2.00$ the calculated temperature increase $\Delta T$, considering the variation of $\sigma_{abs}(\lambda_{pump})$, $S^{-1}$, and $a_0$, is 222 K and 265 K, respectively. These estimates of $\Delta T$, indicated by black arrows in Figs. 2(d-f), are in good agreement with the temperature scale based on the variation of $\Delta H_c(P,T)$. This equivalence between the thermometric scales obtained from Eq. (1) and the micromagnetic simulations indicates that efficient, robust, and reliable temperature control by optical means is possible in hybrid plasmonic-magnetic systems.

Until now, we considered a pump beam with light polarization parallel to the nanoislands' major axis only, which is associated with the strongest plasmonic excitation in the system. In contrast, a pump beam with the same wavelength but polarized along the minor ellipse axis is expected to have little effect, since $\sigma_{abs}(\lambda_{pump})$ is more than ten times smaller in this case, see Figs. 1(h-k). As an example, for islands with $AR = 1.75$ and at wavelength of 785 nm, the simulated absorption cross sections $\sigma_{abs}$ for light polarized along the major (minor) axis differ by more than a magnitude, with values of $\sigma_{abs}(\updownarrow)=5.0\cdot10^4$ nm$^2$ ($\sigma_{abs}(\leftrightarrow)=0.3\cdot10^4$ nm$^2$), as highlighted by open circles in Fig. 1(i). Consequently, thermoplasmonic heating and its effect on the magnetic properties is much less pronounced, as demonstrated in Fig. 3 with the comparison of MOKE hysteresis loops proportional to $M(H, 0)$ and $M(H, P_{pump})$ measured with different incident light polarization while leaving the other experimental parameters unchanged. If the light polarization is parallel to the ellipse major axis, Fig. 3(c), the coercive field decreases by $\Delta H_c(\updownarrow) \approx -46\%$ for a maximum power of $P_{pump} = 60$ mW (red line). In contrast, for light polarization along the ellipse minor axis, Fig. 3(d), the two hysteresis loops are almost the same, and only a minor reduction of the coercive-field in the order of $\Delta H_c(\leftrightarrow) \approx 3\%$ is detected (blue line). This large difference in response allows to use the light polarization as an additional optical degree of freedom, apart from pump power, to control the thermoplasmonic heating.

Based on these encouraging results of efficient and selective optical heating on isolated nanoislands, we now turn towards more complex assemblies of close-packed nanoislands of different orientations. Such structures, like 4-vertices of two pairs of perpendicular nanoislands, schematically shown in Fig. 3(d), are used as basic building blocks for artificial square ice or nanomagnetic logic circuits.[5,24] In this work, 4-vertices with nanoislands equivalent to the aforementioned samples with $AR = 1.75$ were fabricated, as schematically shown in Fig. 3(d). SEM images, experimental extinction spectra as well as simulated absorption cross sections are shown in Fig. S2. The optical response of the vertices is largely determined by the plasmonic excitations of the individual horizontal and vertical islands, and minor shifts of the resonance peaks occur due to optical near-field coupling between neighbouring elements.

The MOKE hysteresis loops of 4-vertices mainly probe the magnetic response of one sublattice only, namely the switching of the nanomagnets that have their major (magnetic easy) axis parallel to the applied field $H$ (black islands in Fig. 3(d)), while the background slope is associated with the magnetic response of those islands with the field applied along the minor (magnetic hard) ellipse axis (marked gray in Fig. 3(d)). In contrast, the choice of pump beam polarisation determines in which of the two sub-lattices plasmons are excited. In Fig. 3(e,f) hysteresis loops obtained under ambient conditions ($P_{pump} = 0$) are shown as thick black lines, whereas red and blue thin lines denote hysteresis loops measured for a pump beam ($P_{pump} \neq 0$) polarized along the vertical and horizontal direction, respectively. The amount of sub-lattice-selective heating depends on the beam diameter $w_{FWHM}^{pump}$, which in Eq. (1) regulates the balance between the localised self-heating term $\Delta T_{self}$ and the effect of thermal diffusion $\Delta T_{coll}$, respectively the equilibriation due to collective heating effects. As shown in Fig. 3(e), for a beam with $w_{FWHM}^{pump} \approx 15$ μm, the coercive field variation is about $\Delta H_c(P_{pump}) \approx -15\%$ for both light polarisations, and the almost equivalent response indicates that $\Delta T_{coll}$ dominates in Eq. (1).[37] In contrast, pumping with a beam with $w_{FWHM}^{pump} \approx 5$ μm, the reduction of the coercive field $\Delta H_c(\updownarrow) = -26\%$ with light polarised in the vertical direction (red line) is larger than the reduction $\Delta H_c(\leftrightarrow) = -15\%$ for horizontal light polarisation (blue line), and sublattice-specific heating is partially restored.

As our experimental results indicate, selective heating of perpendicular sublattices can be achieved under specific circumstances: One strategy to avoid collective plasmonic heating is to consider well-separated elements only, as shown in Figs. 3(a-c). However, for the use in artificial spin systems or nanomagnetic computation circuits, strong magnetostatic interactions are required, neccessitating small inter-element distances to ensure proper operation. Therefore, another stategy to achieve selective heating is to excite less nanoplasmonic heaters simultaneously by focusing the pump beam, *i.e.* reducing the beam diameter $w_{\text{FWHM}}^{\text{pump}}$, and thus reduce the effect of the contribution $\Delta T_{\text{coll}}$ in Eq. (1). The usefulness of this approach, however, is limited by the challenge to achieve sub-micron focusing and to reliably position the light beam on selected vertices in nanomagnetic arrays.

To further illustrate the relationship between the pump beam diameter and the selective heating, we simulated the thermal properties of a 7x7 4-vertex array under illumination of a vertically-polarized beam tuned to the major-axis plasmonic resonance (see Methods), as shown in Fig. 4(a-c): Here, for three different values of $w_{\text{FWHM}}^{\text{pump}}$, the time-dependent temperature increase of the central vertical (blue dots) and horizontal islands (red circles) are shown, for a pump beam switched on (off) at $t = 0$ ($t = 5$ µs). In all three cases, the vertical islands, *i.e.* those parallel to the beam polarization and thus heated more efficiently by the plasmonic excitation, reaching a higher temperature than the horizontal islands.

Reduction of the beam diameter $w_{\text{FWHM}}^{\text{pump}}$ decreases the spatial extent of the illuminated region and thus the number of nanoantenna simultaneously excited, and the temperature difference between the perpendicular sublattices increases. The amount of selective heating by the simultaneous excitation of two distinct plasmonic sublattices with different absorption cross sections $\sigma_{\text{abs}}^{\parallel}$ and $\sigma_{\text{abs}}^{\perp}$ parallel and perpendicular to the incident light polarisation can be expressed by the following term:

$$sel = \frac{\Delta T^{\parallel} - \Delta T^{\perp}}{\Delta T^{\parallel}}. \quad (2)$$

Here, $\Delta T^{\parallel} = \Delta T_{\text{self}}(\sigma_{\text{abs}}^{\parallel}) + \Delta T_{\text{coll}}(\sigma_{\text{abs}}^{\parallel}) + \Delta T_{\text{coll}}(\sigma_{\text{abs}}^{\perp})$ and $\Delta T^{\perp} = \Delta T_{\text{self}}(\sigma_{\text{abs}}^{\perp}) + \Delta T_{\text{coll}}(\sigma_{\text{abs}}^{\perp}) + \Delta T_{\text{coll}}(\sigma_{\text{abs}}^{\parallel})$ denote the temperature increase of the different sublattices due to element-specific heating and the collective heat transfer according to Eq. (1). Values for *sel* range from 0% in the case that both sublattices equilibrate to the same temperature under steady-state illumination, *i.e.* $\Delta T^{\parallel} = \Delta T^{\perp}$, to 100% for perfect sublattice-specific excitation with $\Delta T^{\perp} = 0$. The sub-lattice specific heating for the cases shown in Figs. 4(a-c), as described by Eq. (2), gives values of *sel*=23% for $w_{\text{FWHM}}^{\text{pump}}$ =5.0 µm, *sel*=29% for $w_{\text{FWHM}}^{\text{pump}}$ =2.5 µm, and *sel*=43% for $w_{\text{FWHM}}^{\text{pump}}$ =1.0 µm. Snapshots of thermal maps taken at $t = 5$ µs (*i.e.* just before the pump beam is switched off), Figs. 3(a-c), illustrate how the heat is diffusively re-distributed via the substrate, leading to an increase of the global background temperature and a sizable thermal gradient. These simulation results agree with the mechanism outlined above and the prediction of Eq. (1) for the central temperature increase $\Delta T$.

As shown by the time-dependent temperature evolution curves in Figs. 4(a-c), under constant illumination the system requires up to a few microseconds to reach its equilibrium temperature distribution. In contrast, the temperature *difference* between the two sublattices (shaded gray area) attains its final value within the time step of the simulation. This disparity in timescales stems from the fact that the underlying physical mechanisms are different: Plasmonic excitations lead to a fast temperature rise within a few picoseconds governed by electron-phonon coupling in the nanoscale volume of the antenna.[28,31] In contrast, reaching the steady-state equilibrium of the entire system is a slow process with rise times of tenths to several microseconds, limited by material properties such as density, heat capacity, and thermal conductivity that govern the dynamics of thermal diffusion via the quasi-infinite substrate.[31,36]

Due to the fast excitation of plasmonic oscillations within the nanoantenna, as compared to the slow temperature increase of a larger array, full selective heating of one sub-lattice can be restored, even when using a non-focussed beam, under pulsed illumination using a short pulse on a time scale where thermal diffusion is not relevant. Such behavior is indeed reproduced by time-dependent thermal simulations using a Gaussian pulse with $\tau_{\text{FWHM}}^{\text{pump}}$ =2 ns, Fig. 4(d), and $w_{\text{FWHM}}^{\text{pump}}$ =30 μm to excite the system (see Methods), as shown in Fig. 4(e,f): The temperature of the vertical islands (blue dots) increases dramatically, whereas the horizontal islands (red circles) remain virtually unaffected. This leads to a large selective heating, with a value of $sel \approx 95\%$ within the first 7 ns, and a large temperature offset, gray area in Fig. 4(f), over a time of approximately 10 ns. The snapshots of the thermal distribution at different times, Fig. 4(g), illustrate that the temperature increase of the vertical islands is uniform throughout the sample as the pump beam diameter is much larger than the illuminated array. Furthermore, and in contrast to the larger-time-scale simulations in Fig. 4(a-c), the substrate temperature remains largely at ambient conditions. Thus, perfect sublattice-selective heating on extended arrays of nanoislands can be achieved on time scales from a few picoseconds (limited from below by the speed of the plasmonic excitation) up to nano- to microseconds (limited from above by the time constant of thermal diffusion). As moment reorientations of the magnetic nanoislands also occur on these time scales, such thermoplasmonic heating schemes therefore lend itself to achieve fast thermal equilibration in interacting nanomagnetic samples.

DISCUSSION

Our experimental and simulation results demonstrate the possibility to control the magnetic properties in hybrid plasmonic-magnetic nanostructures, such as saturation magnetization, coercive fields, and reversal dynamics, by optical means. This approach therefore enables thermal control beyond global heating, *i.e.* by thermal contact to a substrate, allowing for spatially-selective (by beam focus), element-specific (by beam polarization) and fast (by pulsed illumination) heating in densely-packed magnetostatically-coupled arrays of nanomagnets. Due to the composite nature of the permalloy-gold building blocks, in the following we wish to discuss design considerations to facilitate efficient light-matter interaction and to tailor the magnetic properties required for strongly-interacting fluctuating artificial spin systems.
To achieve effective and selective photo-induced nanoscale heating, narrow plasmonic resonances with a large absorption cross section must be efficiently excited. This necessitates the use of a hybrid nanostructure, as a control experiment on permalloy-only elements, see Fig. S3, demonstrates: Here, the extinction spectrum shows a weak variation with wavelength only, and no pronounced plasmonic resonances, Fig. S3(a), such that neither strong nor direction-dependent thermoplasmonic heating is observed in power-dependent MOKE measurements, Fig. S3(b). Hybrid structures combining magnetic and plasmonic constituents are therefore indispensable to achieve the desired functionality. The light-matter interaction at the desired pump wavelengths required for efficient thermoplasmonic heating can be maximized by sample design, *i.e.* the size and shape of the gold nanoantenna and optical and thermal properties of the substrate material. Suitable sample dimensions and rough estimates for the final temperature can be obtained by combining simulations of the absorption cross section $\sigma_{\text{abs}}(\lambda, \boldsymbol{E})$ and Eq. (1),[31] and the plasmonic properties of fabricated samples can be conveniently tested using optical spectroscopy. Therefore, the robust, reliable, and easily quantifiable behavior of plasmonic nanostructures lends itself for convenient heating of nanomagnets.

Large equilibrium temperature increases by thermoplasmonic heating, as estimated by Eq. (1), can be easily achieved with plasmonic heating by varying the pump power $P_{\text{pump}}$, the beam diameter $w_{\text{FWHM}}^{\text{pump}}$ or by using short laser pulses. If selective heating of sublattices of differently-oriented magnets is desired, illumination conditions such as beam diameter and pulse duration are relevant: Especially the simultaneous excitation of many plasmonic elements and thermal diffusion can lead to a sizable increase of the global background temperature, which in turn diminishes the effect of selective heating. This undesirable effect can be mitigated by focusing the pump beam (*i.e.* reducing the beam diameter $w_{\text{FWHM}}^{\text{pump}}$) or by illuminating small-scale structures with a handful of individual elements only. Also, as the timescale of thermal diffusion is in the order of several tens to hundreds of nanoseconds, short laser pulses can be used to selectively excite specific sublattices.

Laser pulses with a duration of a few ten nanoseconds down to several ten picoseconds also match the expected time scale of magnetic relaxation of nanomagnetic elements at elevated temperatures: The fluctuation frequency $\nu$ of the thermally-activated magnetization reversal in individual nanomagnets can be described by an Arrhenius law $\nu = \nu_0 e^{-\frac{E_K}{k_B T}}$, with the anisotropy energy barrier $E_K$, the temperature $T$, and the Boltzmann constant $k_B$. The so-called attempt frequency $\nu_0$ is in the order of $10^{9...12}$ Hz, depending on material parameters, particle size and shape as well as temperature.[38-40] Usual values for the anisotropy energy barrier $E_K = E_{\text{island}} + E_{\text{dip}}$ are in the order of meV to eV,[5,17,24] and depend on the single-island shape anisotropy $E_{\text{island}}$ (due to material parameters and nanomagnet size and shape, *e.g.* quantified by the aspect ratio $AR$) as well as magnetostatic interactions $E_{\text{dip}}$ (due to the presence of neighbouring elements). Typical timescales $\nu$ for the reversal dynamics of nanomagnets at elevated temperatures therefore range from pico- to nanoseconds, matching the time window for which sublattice-selective heating via pulsed illumination is possible. Thus, thermoplasmonic excitations with pico- to nanosecond-long laser pulses allows for fast, selective, and efficient thermalization protocols even for arrays containing thousands of densely-packed nanomagnetic elements.

Contrary effects of excessive plasmonic heating, caused by too high pump power, are light-induced structural changes of the hybrid gold-permalloy nanostructures. Especially alloy intermixing[41,42] or chemical modifications of the permalloy component[43,44] could irrevocably change the plasmonic and magnetic properties, rendering the functionality of the system void. In addition to potential optical damage, too high temperatures are also diametral to strong magnetostatic interactions, as the contactless dipolar coupling via stray fields is proportional to the product of the neighboring islands' magnetization, and decreases like $E_{\text{dip}} \propto [M_S(T)]^2$ as the saturation magnetization $M_S(T)$ is reduced at elevated temperatures, and vanishes above the magnetic Curie temperature $T_C$ of the material.[45] Fortunately, to achieve efficient thermal equilibration in strongly-coupled nanomagnetic arrays, the kinetics governed by the Arrhenius equation are more relevant than the energetics, and even intermediate temperature increases will lead to a significant rise of the frequency of thermally-assisted moment reorientations while magnetostatic interactions remain strong. Choosing the right size, shape, and arrangement of hybrid gold-permalloy nanoislands therefore allows the targeted design of nanomagnetic arrays with desired relaxation behavior for artificial spin systems or nanomagnetic computation. In particular, the spatial extension of nanomagnetic logic circuits, containing a handful elements only, is on the length scale of 1 μm,[3,24] and thus the targeted heating of single gates is conceivable with a focused beam.[46] Furthermore, the relaxation behavior of a nanomagnetic gate can be drastically altered by heating a selected subset of elements only, by changing the light polarization, offering yet another degree of freedom for reconfigurable nanomagnetic computation.

CONCLUSIONS & OUTLOOK

In this work, we demonstrate the use of photo-induced thermoplasmonic heating to control the thermally-activated magnetization reversal in arrays of nanomagnets, using hybrid nanoscale Au-Py elements for the desired optical, thermal, and magnetic properties. We maximized the light-matter interaction at technologically-relevant visible/near-infrared wavelengths by choosing a suitable island aspect ratio and demonstrated the optical control of the magnetic coercive field by the power and polarization of the pump beam. For 4-vertex structures with inequivalent sub-lattices we explored strategies to achieve element-selective heating, which facilitate thermalization protocols that are impossible to achieve by conventional global heating.

Light-controlled heating schemes are quite simple to implement under different experimental conditions, from table-top setups to large-scale facility measurements, and offer additional advantages, such as reduced thermal drift. Using this approach for artificial spin systems, thermalization routines can be tested for many systems on one chip, drastically simplifying current approaches that require physically separate samples for each heating cycle. In addition, heating protocols with varying pulse position, intensity, polarization, and duration allow for step-by-step nanomagnetic computation at GHz frequencies. In the future, we expect element-sensitive and fast plasmon-assisted photo-heating of nanomagnets to be employed for the investigation of equilibrium properties and exotic thermal excitations in artificial spin ices, such as magnetic monopoles,[14,47] as well as flexible thermalization protocols for reconfigurable nanomagnetic logic,[3,24,46,48,49] where individual computational units are addressed individually by scanning a focused beam over the region of interest.

METHODS

**Sample fabrication.** Nanopatterned samples were fabricated with electron-beam lithography on transparent Pyrex glass substrates, using a standard lift-off procedure. Before spin-coating the resist (ZEP520A-7 by ZEON), the substrate was coated with a 4 nm-thick titanium film to avoid charging effects and to improve adhesion. After electron-beam exposure and development (30 s in ZED-N50 (ZEON), and 30 s in isopropanol to stop the reaction) a tri-layer Au(25 nm)|Py(10 nm)|Au(5 nm) film was deposited at a rate of 0.5 Å/s using electron-beam evaporation at a base pressure of approximately $7\cdot10^{-6}$ mbar. After deposition of the metal layers, the resist was dissolved in ZDMAC (ZEON), lifting off unwanted material, such that only the patterned nanoislands remain. Throughout this work, elliptical nanoislands with a minor axis fixed to $a_{\text{minor}} = 100$ nm, and different aspect ratio $AR > 1$, *i.e.* major axis length $a_{\text{major}} = AR \cdot a_{\text{minor}}$, have been considered.

**Optical characterization.** The wavelength-dependent transmittance of linearly polarized light, ***E*** parallel to $a_{\text{minor}}$ or to $a_{\text{major}}$, $T(\lambda, \boldsymbol{E})$ was measured for all patterned arrays in the spectral regime of 400 nm to 1100 nm, using commercial spectrometers by StellarWiz and Ocean Optics. The spectra $T(\lambda, \boldsymbol{E})$ have been normalized to the transmittance of the titanium-coated glass substrate, which was measured on the same substrate in a region without nanopatterned arrays.

**COMSOL simulations of optical properties.** Optical absorption cross sections for the fabricated elliptical nanoislands, as well as four-island vertices, for ***E*** parallel to the minor and major ellipse axes have been calculated using COMSOL Multiphysics simulations ("RF" module).[34] The refractive indices for gold[50] and permalloy[51] have been obtained via Ref. (52).

**Probing of thermoplasmonic heating upon magnetic hysteresis loops.** To characterize the influence of plasmonic heating on the magnetic properties of the samples, a dedicated experimental setup has been built, Fig. S3, allowing to optically pump the gold nanostructures from the back of the substrate while simultaneously measuring the magneto-optical Kerr effect (MOKE) in reflection geometry from the front of the sample. After impinging on the sample with an incidence angle of about 35°, the reflected probe beam passes through a photoelastic modulator and a polarizer, before reaching the photodetector. The use of a lock-in amplifier, locked to the detected signal at the 50 kHz modulation of the beam polarization, allows highly-sensitive measurement of the Kerr ellipticity.[53] In the considered longitudinal MOKE geometry, the measured signal is proportional to the component of the sample magnetization $M(H)$ lying along the intersection between the sample surface and the light scattering plane. The magnetic field $H$ is applied along the same direction. The MOKE probe laser had a wavelength of $\lambda_{probe}$ =532 nm. Due to the small extinction at this wavelength, no efficient thermoplasmonic heating is expected to occur, and we experimentally verified the negligible effect of heating on the magnetic hysteresis loops with varying intensity and polarization of the probe beam. On the other hand, the linearly-polarized pump laser had a wavelength of $\lambda_{pump}$ =785 nm to match the wavelength where a maximum plasmonic heating efficiency and selectivity was expected. The pump beam had a full-width at half maximum (FWHM) diameter of $w_{FWHM}^{pump}$=25 μm and an adjustable power $P_{pump}$ from 0 to 100 mW. The MOKE probe beam had a diameter of $w_{FWHM}^{probe}$ ≤15 μm, overlapping the central region of the pump beam where a uniform temperature increase is expected. MOKE hysteresis loops were measured with varying pumping conditions, using field steps of ~10 Oe. The saturation signal of the hysteresis loops taken under ambient conditions ($P_{pump}$ =0, $T$ =300 K) were normalized to one. The coercive field was determined as the mean between the zero-crossing on up- and down-sweeps of the field: $H_c = \frac{1}{2}(H_{c\nearrow} - H_{c\searrow})$.

**Micromagnetic simulations with OOMMF.** The temperature-dependent variation of the coercive field was simulated using the three-dimensional Object Oriented Micromagnetic Framework (OOMMF) software.[54] The elliptical nanoislands with a minor axis of $a_{minor}$ =100 nm and aspect ratios $AR$ =1.50, 1.75, and 2.00 were divided into a regular mesh of cubic cells with dimensions 2.5×2.5×2.5 nm³. An exchange constant of $A_{ex}$ =1.3·10⁻¹¹ J/m, vanishing magnetocrystalline anisotropy, a damping constant $\alpha$ =0.05, and a temperature-dependent saturation magnetization $M_s(T) = M_s^0 \cdot \left(1 - \frac{T}{T_C}\right)^\beta$ have been considered. A saturation magnetization of $M_s^0$=6·10⁵ A/m and a Curie temperature $T_C$ =843 K were used, and the exponent $\beta$ =0.35 was obtained by a fit to the temperature-dependent magnetization data presented in Ref. (9).

**COMSOL simulations of thermoplasmonic properties.** To elucidate the effect of different illumination conditions on the plasmonic heating of densely-packed nanostructured, we performed thermal simulations under steady-state and pulsed illumination of a finite array of 7x7 vertices, with a lattice constant of 680 nm and a vertex gap of $d_{vertex}$ =100 nm, using ellipse islands with a minor axis of $a_{minor}$ =100 nm and an aspect ratio of $AR$ =1.75. The simulations were implemented using the COMSOL Multiphysics "Heat Transfer in Solids" module.[34] For the Pyrex glass substrate, a thermal conductivity of $\kappa_{sub}$ =1.4 W m⁻¹ K⁻¹ has been considered, and the surrounding medium was assumed to be vacuum with $\kappa_{env}$ = 0.[55] The mass density and specific heat of the nanoellipses have been calculated as a volume average of the characteristic values for the 30 nm thick gold (*i.e.* 25 nm + 5 nm) and 10 nm thick nickel layer.[56] For the simulations to reach the steady-state under constant illumination the pump beam power $P$ was chosen such that the maximum temperature reached in the central vertical island remains fixed to ~200 K upon variation of the beam diameter $w_{FWHM}^{pump}$. For heating dynamics, where a Gaussian light pulse of duration $\tau_{FWHM}^{pump}$ =2 ns excites the system, $w_{FWHM}^{pump}$ was fixed to 30 μm and the beam power $P_{pump}$ was adjusted to deliver the desired amounts of energy of ≈5 nJ to the system, which generates a maximum temperature increase of ≈200 K. The values for the optical absorption $\sigma_{abs}(\lambda, \boldsymbol{E})$ at $\lambda$=808 nm was taken from the calculated absorption cross section spectra for the vertex structures, as shown in Fig. S2(c). The baseline temperature of all simulations was $T_0$ =300 K.

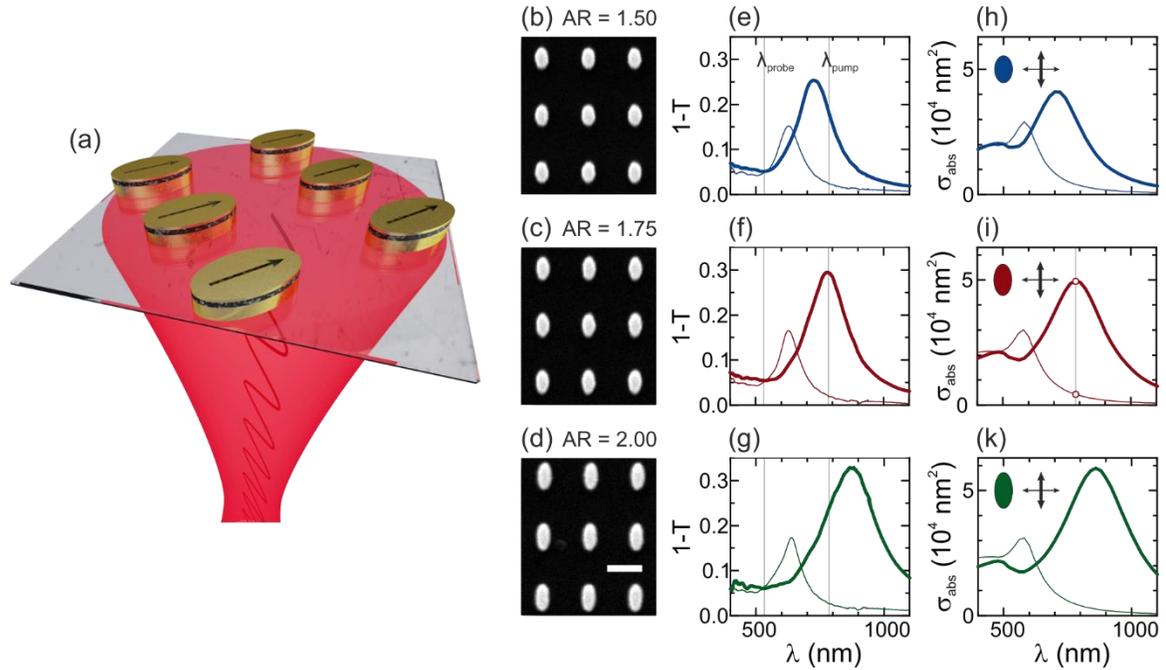

**Fig. 1. Variation of plasmonic resonances with ellipse aspect ratio.** (a) Schematic rendering of hybrid gold-permalloy nanoislands, illuminated with a polarized laser beam from the backside of the glass substrate. (b-d) Scanning electron micrographs of tri-layer Au-Py-Au ellipses with varying aspect ratio $AR$. Scale bar in (d) measures 250 nm. (e-g) The measured extinction spectra $1 - T(\lambda, \boldsymbol{E})$ for visible and near-infrared light, with thick (thin) lines denoting spectra taken with light polarization along the major (minor) ellipse axes. An increase in the ellipses' aspect ratio $AR$ leads to a red shift and enhanced light-matter interaction for the plasmonic resonance (thick lines). In contrast, the minor-axis resonance around 630 nm (thin line) remains unaffected. Wavelengths of the MOKE probe beam, $\lambda_{\text{probe}}$, and plasmonic pump beam, $\lambda_{\text{pump}}$, are indicated by gray lines. (h-k) Absorption cross sections $\sigma_{\text{abs}}(\lambda, \boldsymbol{E})$, simulated with COMSOL, for light polarization along the ellipse major and minor axes (thick and thin lines, respectively), indicate that strong thermoplasmonic heating is expected for the plasmonic resonances. The open circles in (i) show the pronounced polarization-dependent values of the absorption cross section.

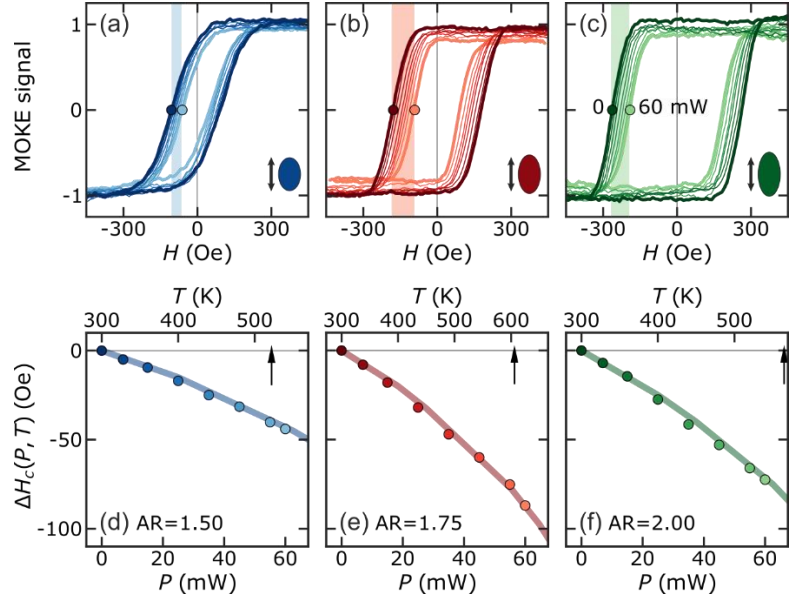

**Fig. 2. Influence of thermoplasmonic heating on magnetic properties.** (a-c) Power-dependent MOKE hysteresis loops measured for different aspect-ratio islands, pumped with a beam of wavelength $\lambda_{\text{pump}} = 785$ nm, polarization along the ellipses' major axis, and varying the power $P_{\text{pump}}$. The large circles indicate the coercive field at ambient conditions (which increases with the nanoislands aspect ratio), and at the maximum used power of 60 mW. The largest reduction of coercive field $\Delta H_c(P_{\text{pump}})$ is observed for the sample with $AR = 1.75$, which shows the strongest light-matter interaction $\sigma_{\text{abs}}(\lambda_{\text{pump}}, \boldsymbol{E})$ at the pump wavelength. (d-f) Power-dependent decrease of coercive field $\Delta H_c(P_{\text{pump}})$ from MOKE measurements (filled dots, bottom scale), compared to the temperature-dependent decrease of the coercive field $\Delta H_c(T)$ from micromagnetic OOMMF simulations (solid line, top scale). The maximum temperature increases obtained from magnetic simulations agree reasonably well with those estimated from Eq. (1) for $P_{\text{pump}} = 60$ mW, indicated by arrows, which consider the optical properties of the nanoislands and illumination conditions only.

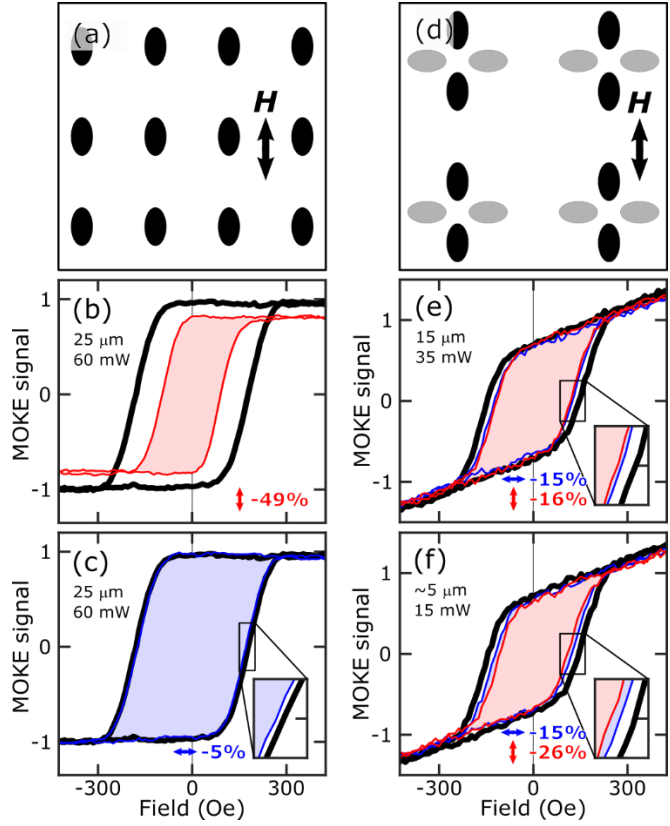

**Fig. 3. Selective control of magnetic hysteresis via light polarization.** (a,d) Schematic drawing of (a) single-island arrays and (d) arrays of 4-vertices made from nanoislands with aspect ratio $AR = 1.75$. The MOKE signal mainly probes the magnetic response of the vertical islands (black) whose long, magnetic easy, axis is parallel to the applied field $H$. (b,c,e,f) Hysteresis loops measured under ambient conditions ($P_{\text{pump}} = 0$, black thick line) and with maximum pump beam at wavelength $\lambda_{\text{pump}} = 785$ nm polarized along the vertical (red lines) and horizontal (blue lines) direction. (b) For arrays of vertically-arranged ellipses with aspect ratio $AR = 1.75$, a pronounced reduction of the coercive field of -49% is observed when the light polarization is along the ellipse major axis. (c) In contrast, illumination with a horizontal light polarization leaves the hysteresis loops largely unchanged, with a decrease of -5%, indicating that heat generation by plasmonic resonances is negligible in this configuration. (e) For 4-vertex arrays, the MOKE hysteresis loops under vertically-polarized (red) and horizontally-polarized (blue) pump beams are very similar. Even though the polarized pump beam excites plasmons in a selected subset of elements only, *i.e.* either black or gray islands in (c), the generated heat is diffusively re-distributed, leading to a global temperature rise. (f) Sublattice-selective plasmonic heating of the 4-vertices is partially recovered when the beam diameter $w_{\text{FWHM}}^{\text{pump}}$ is reduced from 15 μm to ~5 μm. The *sel* values in (e) and (f) have been estimated according to Eqs. (1) and (2).

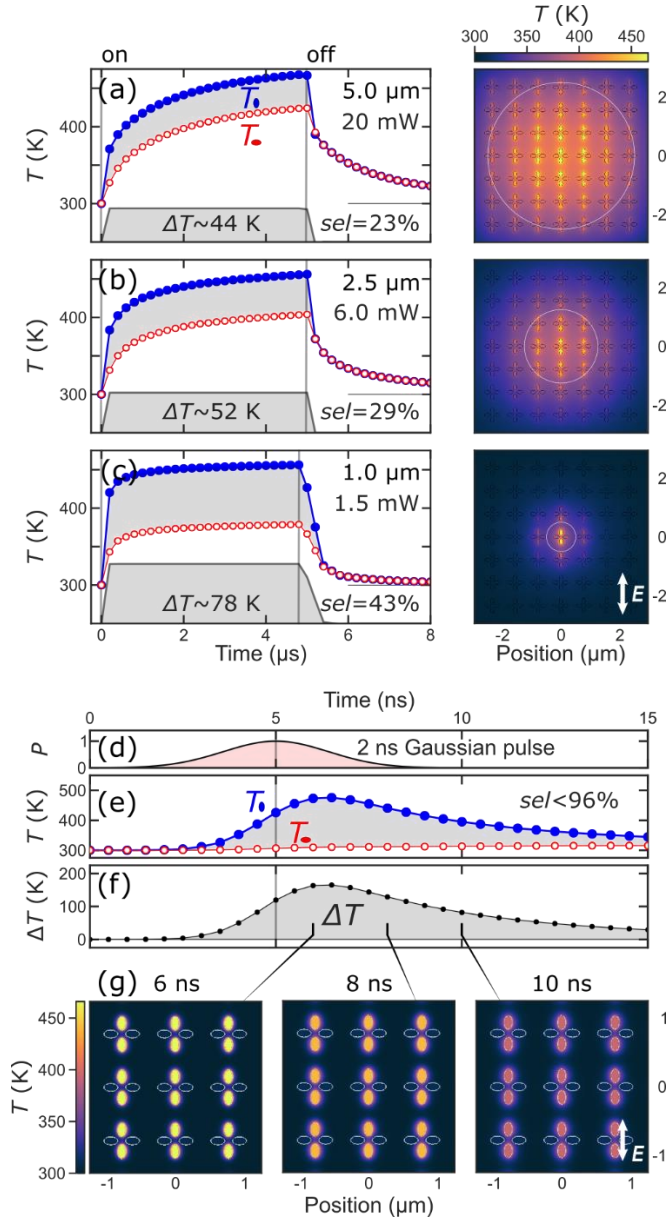

**Fig. 4. Selective heating under different illumination conditions.** (a-c) Time evolution of the temperatures of the central vertical (blue dots) and central horizontal (red open circles) nanoislands under constant illumination which is turned on and off at 0 μs and 5 μs, respectively, (left) and a thermal map at 5 μs (right). For the simulations, a 7x7 array of 4-vertices illuminated with a variable beam diameter $w_{FHWM}^{pump}$ of (a) 5 μm, (b) 2.5 μm, and (c) 1.0 μm (white outline) has been considered. In each case, the power of the beam has been chosen to achieve a comparable maximum temperature increase of ~200 K in the central vertical nanoisland. Although the beam is polarized along the vertical direction, thermal diffusion leads to a sizable heating of the horizontal sublattice and the substrate as well. The temperature difference between vertical and horizontal islands (gray area) increases if less nanoislands are illuminated, *i.e.* for smaller $w_{FWHM}^{pump}$. (d-g) Using a 2 ns short pump pulse ($w_{FWHM}^{pump}$ =30 μm) to excite the plasmonic resonances, the temperature of the vertical islands (blue dots) rises substantially, whereas only little heat is transferred to the horizontal islands (red open circles). (g) As indicated by thermal maps taken at different times. Thus, almost perfect sublattice selectivity can be achieved at ps to ns timescales relevant for magnetic relaxation even for fully-illuminated densely-packed arrays of nanomagnets.

ACKNOWLEDGEMENTS

This work was supported by the Spanish Ministry of Economy, Industry and Competitiveness under the Maria de Maeztu Units of Excellence Programme - MDM-2016-0618 and the Project FIS2015-64519-R, as well as from the European Union under the Project H2020 FETOPEN-2016-2017 "FEMTOTERABYTE" (Project n. 737093).

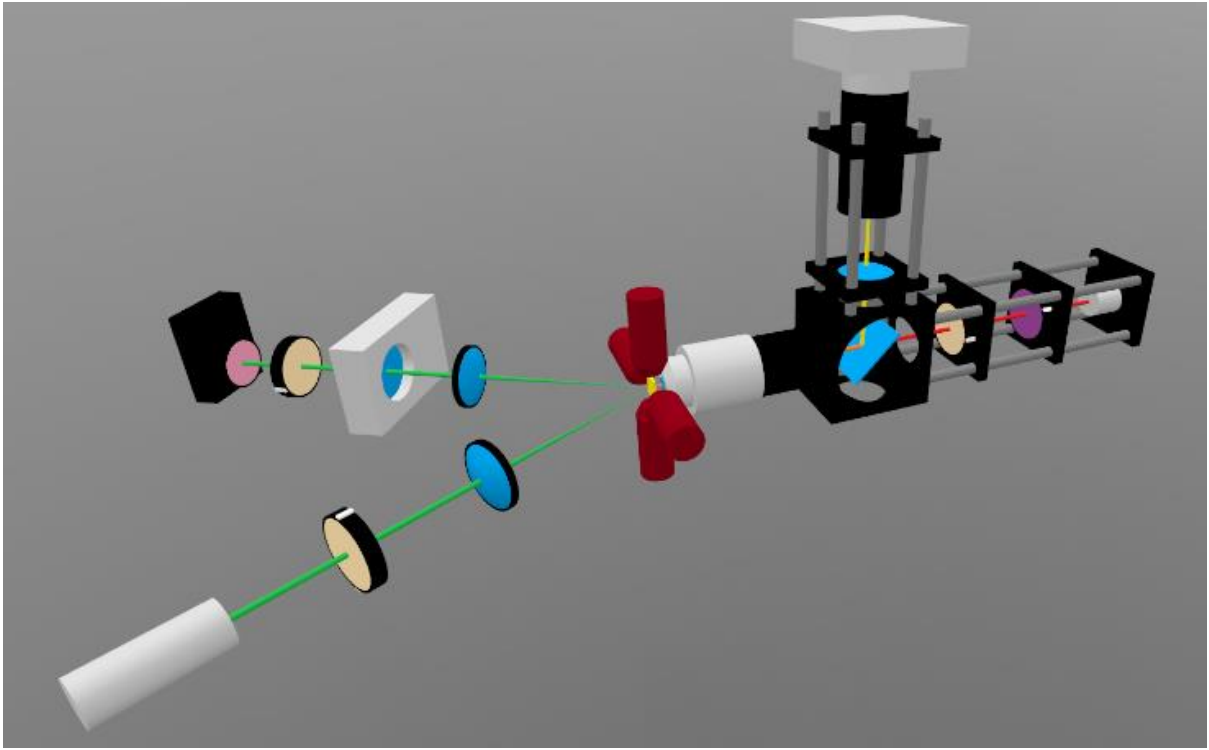

**Fig. S1: Experimental setup.** On the left, the MOKE setup (green light path), which probes the samples magnetic response from the front of the substrate, is shown. The sample is situated between magnetic pole shoes (red, middle). The thermoplasmonic excitation with varying light polarization, power, and focus is obtained from the back of the sample (red light path), with the setup shown on the right side of the illustration. A CCD camera allows checking the spatial overlap of the MOKE probe beam coming from the front of the sample (green) and the plasmonic pump beam coming from the back of the sample (red) on the transparent glass sample substrate.

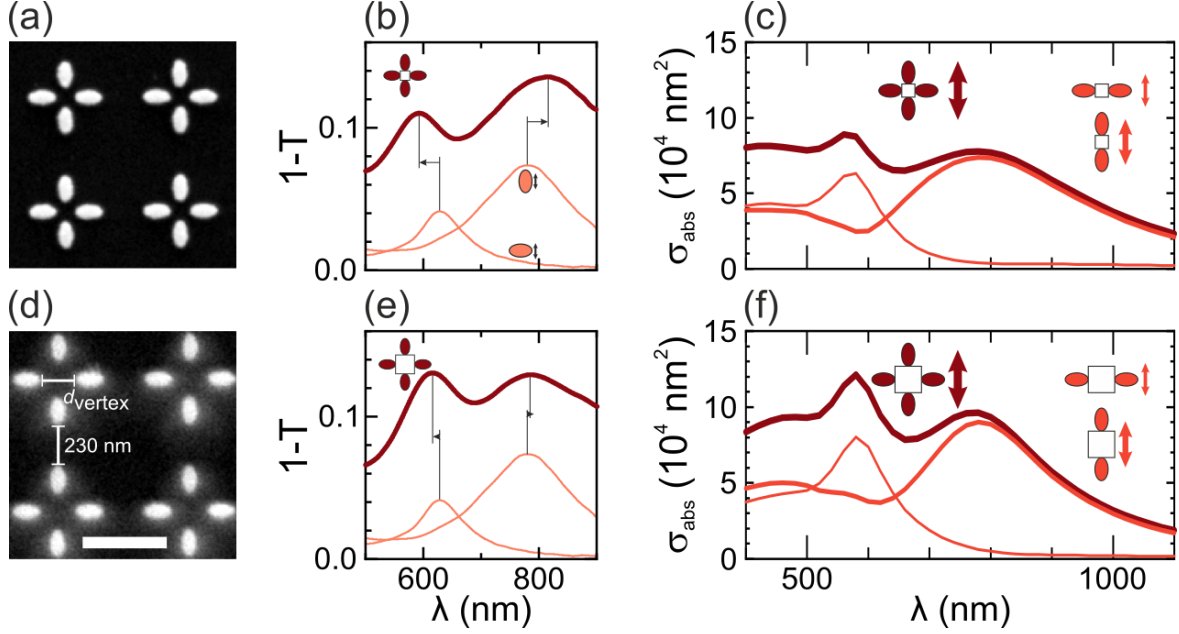

**Fig. S2: Optical properties of 4-vertices.** (a,d) SEM micrographs of 4-vertices with perpendicular islands of aspect ratio $AR =1.75$ ($a_{\text{minor}} =100$ nm) and a center edge-to-edge islands distance of (a) $d_{\text{vertex}} =100$ nm, and (d) $d_{\text{vertex}} =200$ nm. The 4-vertex units are separated by a distance of 230 nm to consider them optically and magnetically isolated (*i.e.* near-field decoupled) from each other. The scale bar in (d) measures 500 µm. (b,e) Experimentally-measured optical extinction spectra $1 - T(\lambda, \boldsymbol{E})$ show two peaks associated with the two plasmonic resonances along the minor and major ellipse axes of the perpendicular islands of the 4-vertices. Near-field coupling of neighboring islands leads to blue- and redshifts of the plasmon peaks, as indicated. For comparison, the single-island spectra, from Fig. 1(f) scaled by a factor of 0.25, are shown as thin lines (due to the smaller particle surface density $S^{-1}$ the extinction is lower for the vertex arrays than the samples shown in Fig. 1). (c,f) Simulated absorption cross sections $\sigma_{\text{abs}}(\lambda, \boldsymbol{E})$ for the horizontal pair (thick light red line), vertical pair (thin light red line), and full vertex (solid dark red line).

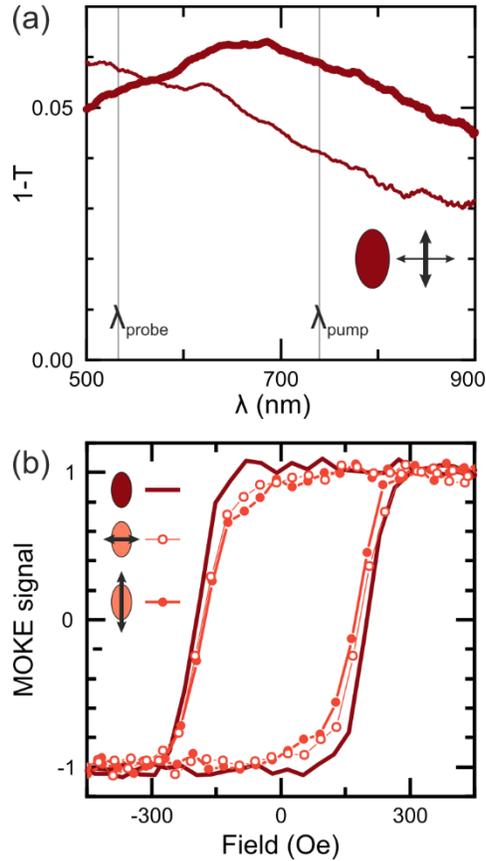

**Fig. S3 Plasmonic and magnetic properties of permalloy-only nanomagnets.** To show the superior thermoplasmonic properties of gold layers, arrays of permalloy-only nanomagnets with an ellipse aspect ratio AR=1.75 ($a_{minor}$ =100 nm, $a_{major}$ =175 nm), have been fabricated. (a) Extinction spectra show a weak wavelength- and polarization dependence, and in general a much weaker light-matter interaction compared to the tri-layer Au-Py-Au islands. The pump wavelength $\lambda_{pump}$ =740 nm (gray line) represent the best compromise between heating efficiency and polarization-dependent selectivity. (b) MOKE hysteresis loops measured under different illumination conditions – no pump (solid line), and a pump beam ($P$ =60 mW) polarized along the ellipse major (closed circles) and minor axis (open circles) – show only a small variation in coercive field ($\Delta H_C$ <25 Oe) and a negligible effect of the light polarization.